\documentclass[sigconf]{acmart}

\AtBeginDocument{%
  \providecommand\BibTeX{{%
    \normalfont B\kern-0.5em{\scshape i\kern-0.25em b}\kern-0.8em\TeX}}}

\copyrightyear{2025}
\acmYear{2025}
\setcopyright{acmlicensed}\acmConference[XX]{xxxxxxxx}{xxx, 2025}{Somewhere on Earth}
\acmBooktitle{XXXXXXX}
\acmDOI{}
\acmISBN{}

\usepackage{multirow}
\usepackage{multicol}
\usepackage{xcolor}
\usepackage{colortbl}
\usepackage{listings} 
\usepackage{natbib}
\usepackage{arydshln} 

\begin{document}


\title[Performant Automatic BLAS Offload on Unified Memory Architecture with OpenMP First-Touch alike Data Movement]{Performant Automatic BLAS Offloading on Unified Memory Architecture with OpenMP First-Touch Style Data Movement}
\author{Junjie Li}
\email{nicejunjie@gmail.com}
\orcid{0000-0002-1051-5927}
\affiliation{%
  \institution{Texas Advanced Computing Center, \\The University of Texas at Austin}
  \streetaddress{10100 Burnet Rd}
  \city{Austin}
  \state{Texas}
  \country{USA}
  \postcode{78758}
} 

\author{Yinzhi Wang}
\email{iwang@tacc.utexas.edu}
\orcid{0000-0001-8505-0223}
\affiliation{%
  \institution{Texas Advanced Computing Center, \\The University of Texas at Austin}
  \streetaddress{10100 Burnet Rd}
  \city{Austin}
  \state{Texas}
  \country{USA}
  \postcode{78758}
}

\author{Hang Liu}
\email{hliu@tacc.utexas.edu}
\orcid{0000-0002-3486-7863}
\affiliation{%
  \institution{Texas Advanced Computing Center, \\The University of Texas at Austin}
  \streetaddress{10100 Burnet Rd}
  \city{Austin}
  \state{Texas}
  \country{USA}
  \postcode{78758}
}

\renewcommand{\shortauthors}{J. Li.}

\begin{abstract}

BLAS is critical for scientific computing. 
While GPUs excel at BLAS operations, porting code to GPUs remains challenging due to code complexity and data transfer overhead. 
Unified memory architectures, like NVIDIA Grace-Hopper, enable cache-coherent memory access between CPU and GPU, addressing conventional bottlenecks. 
Building on previous work for gemm offload, we present SCILIB-Accel, a novel tool for automatic offloading of all level-3 BLAS routines. 
SCILIB-Accel leverages Grace-Hopper’s NVLink C2C interconnect and introduces a Device First-Use policy, inspired by OpenMP’s First-Touch approach, to minimize CPU-GPU transfers. 
Using dynamic binary instrumentation, it intercepts BLAS calls at runtime without code modifications or recompilation. 
Evaluated on quantum physics codes across hundreds of GPU nodes, SCILIB-Accel achieves significant speedups, including a 3× improvement for the LSMS method in the MuST suite on Grace-Hopper versus Grace-Grace. 
SCILIB-Accel is the first tool enabling practical, high-performance automatic BLAS offload for scientific applications.

\end{abstract}




\maketitle

\section{Introduction}
\label{sec:intro}

The Basic Linear Algebra Subprograms (BLAS) serve as a building block for many scientific computing applications and form the foundation for advanced linear algebra libraries such as LAPACK and ScaLAPACK. 
These libraries are extensively used in mathematical software like Mathematica and MATLAB, as well as in data science packages such as NumPy, and in computational chemistry and physics applications. 
Notably, BLAS is heavily utilized in quantum chemistry and quantum physics codes, 
as linear algebra is the natural language of quantum mechanics.

Modern general-purpose Graphics Processing Units (GPUs) are known for their exceptional arithmetic compute power. 
Their raw FP32 and FP64 compute capabilities significantly outpace those of CPUs, 
making GPUs an ideal platform for running BLAS-intensive applications. 
While all major GPU manufacturers provide highly optimized BLAS libraries, 
such as cuBLAS for NVIDIA GPUs and rocBLAS for AMD GPUs,
these GPU libraries have slightly different interfaces and are not drop-in replacements for CPU BLAS libraries.
More critically, using these GPU BLAS libraries, like any other GPU porting task, requires developers to manually manage data movement for optimal performance. 
Consequently, porting large codes, legacy codes and codes with complex workflow to GPU is not trivial, sometimes daunting, and requires significant investment of manpower. 
Furthermore, supercomputers are becoming increasingly GPU-centric due to the rapid advancement of GPUs and the surge of AI applications, 
this trend generates pressing needs to port more scientific codes to GPU, 
but many researchers lack the expertise for GPU porting and face challenges securing funding for pure code-porting efforts.
Given BLAS' central role and its extensive GPU support, there have been numerous attempts to automate GPU usage, 
as outlined in Section \ref{sec:previous-blas-offload}. 
However, limitations inherent to conventional GPU architectures often necessitate frequent data transfers between main memory and GPU memory, resulting in overheads that are unacceptable for practical use. 

In recent years, GPU manufacturers have introduced highly innovative architectures featuring unified memory connected via cache-coherent interconnects, such as AMD's MI250X and MI300X GPUs with Infinity Fabric and NVIDIA's Grace-Hopper with NVLink-C2C. 
Additionally, designs like AMD's MI300A Accelerated Processing Unit (APU) integrate CPUs and GPUs with a single type of memory. 
These innovations eliminate the constraints of conventional architectures and inspire new programming approaches that may make automatic offloading feasible. 

In the previous work \cite{scilib-accel-short, scilib-accel-code}, there was a proof-of-concept framework for symbol interception and replacement using Dynamic Binary Instrumentation (DBI). Preliminary experiments demonstrated the feasibility of automatically offloading *gemm (general matrix multiplication) routines, providing evidence that performant, automatic BLAS offloading is achievable.
In this paper, we extend the implementation to support all level-3 BLAS routines and introduce a novel data management strategy, referred to as the Device First-Use policy. 
Inspired by the OpenMP First-Touch memory management model commonly used in multi-socket CPU or NUMA systems, the Device First-Use policy targets CPU-GPU architectures. 
It operates by migrating data to device memory (e.g., GPU memory) upon its first access by a GPU kernel. This policy is both conceptually simple and practically effective, significantly reducing redundant data transfers in typical BLAS-intensive workloads.
We evaluate the approach across several scientific applications with heavy BLAS usage, scaling up to hundreds of GPU nodes. Results show substantial performance gains. 
Although our experiments were conducted on the NVIDIA Grace-Hopper system, the proposed methodology is broadly applicable to any CPU-GPU system with cache coherency.

The remainder of this paper is organized as follows: 
Section \ref{sec:review} reviews the NVIDIA Grace-Hopper unified memory architectures and related BLAS offloading work. 
Section \ref{sec:implementation} details the implementation of SCILIB-Accel. 
In Section \ref{sec:result}, we apply the tool to two BLAS-intensive scientific computing codes on up to 200 Grace-Hopper nodes and discuss the results along with performance issues of the NVIDIA Grace-Hopper system. 
Finally, Section \ref{sec:conclusion} concludes the paper.

\section{Background and Related Work}
\label{sec:review}
 
\subsection{Coherent memory in NVIDIA Grace-Hopper} 
\label{sec:uma}


\begin{table}[ht]
\centering 
\captionsetup{justification=centering} 
\centering
\caption{STREAM Bandwidth (GB/s) \\ on GH200 (120GB LPDDR5X model) } 
\label{tab:stream}
\vspace{-3mm}
\begin{tabular}{l|c|c c}
\hline
               & & LPDDR5 & HBM3 \\
\hline
\multirow{4}{*}{CPU{\scriptsize} }  
               & Copy  &446.46 &   145.56 \\
               & Mul   &438.58 &   145.50 \\
               & Add   &435.28 &   141.94 \\
               & Triad &418.22 &   141.94 \\
\cline{1-4}
\multirow{4}{*}{GPU}
               & Copy  &464.75 &   3364.55 \\
               & Scale &471.07 &   3364.55 \\
               & Add   &474.59 &   3668.78 \\
               & Triad &473.75 &   3679.50 \\
\hline
\end{tabular}
\end{table}

In conventional architectures, GPU and host memory exist in separate memory space, preventing direct access between CPU and GPU memory. 
To partially address this limitation, CUDA 6 introduced managed memory, 
allowing a single memory address space that is accessible from any GPU or CPU. 
This system operates through implicit page migration triggered by page faults managed by the CUDA runtime.
The NVIDIA Grace-Hopper superchip takes a step further by featuring closely integrated CPU and GPU units along with LPDDR5x and HBM3e memory subsystems, connected by the high-bandwidth and cache-coherent NVLink Chip-2-Chip (C2C) interconnect\cite{gh-whitepaper}. 
While Grace-Hopper maintains compatibility with traditional GPU memory management where GPU memory is exclusively managed by the GPU's memory management unit,
it more importantly implements a single system-managed page table where both CPU and GPU can cache-coherently access all memory subsystems without page movement. 
In this unified memory architecture, the two types of memories appear as two NUMA domains, similar to memories in a two-socket CPU system. 

Although the two memory subsystems can be accessed cache coherently by both the CPU and GPU, 
the bandwidth varies significantly between different access patterns, as shown in Table \ref{tab:stream}. 
When the CPU accesses its local LPDDR5X memory, it achieves a descent bandwidth of more than 400 GB/s. 
The GPU, when accessing its local HBM3 memory, delivers substantially higher performance, reaching 3.6 TB/s. The NVLink-C2C interconnect provides 450 GB/s of bandwidth in each direction, 
which is sufficient to saturate the LPDDR5X bandwidth and allows the GPU to access remote LPDDR5X memory at over 400 GB/s.
In contrast, CPU access to HBM3 memory is considerably slower, with measured bandwidth of approximately 140 GB/s. 
These bandwidth disparities indicate that data locality remains crucial for Grace-Hopper systems. To achieve optimal performance, developers must explicitly manage data placement and movement, similar to conventional GPU architectures, rather than rely solely on coherent unified memory access for production workloads.

\subsection{Previous automatic BLAS offload attempts}
\label{sec:previous-blas-offload}


Numerous attempts have been made to automatically accelerate CPU BLAS calls since the early adoption of GPUs in HPC. 
 Cray LIBSCI\_ACC\cite{cray-libsci-acc, cray-libsci-acc-slide}, available for over a decade, was deployed on the Titan supercomputer for NVIDIA Tesla K20 GPU, supporting selected BLAS, LAPACK, and ScaLAPACK routines for offload when the library module is loaded. 
 Similarly, IBM ESSL\cite{ibm-essl-offload} is capable of automatically offload selected BLAS, LAPACK, and FFTW calls but requires the accelerated version of math library, libesslsmpcuda, linked. 
 NVIDIA's NVBLAS\cite{nvidia-nvblas} serves as a drop-in replacement for CPU BLAS calls, allowing users to configure host BLAS libraries and selected routines for offload.  
 By $LD\_PRELOAD$ NVBLAS, dynamically linked CPU BLAS is replaced without relinking.   
 Unfortunately, the NVBLAS tool is heavily over-engineered, it uses the cuBLASXT as the backend instead of cuBLAS and has an acceptable overhead\cite{scilib-accel-short}.  
 
These tools make offload decisions at runtime based on workload sizes and they handle data movement automatically.  
Overall, these libraries are tailored for conventional GPU architectures, where frequent data movement is unavoidable, therefore they suffer poor performance for small and medium sized matrix math in real workloads.
 
\section{Performant Automatic BLAS Offload} 
\label{sec:implementation}

\begin{figure}[ht]
    \centering
    \includegraphics[width=\linewidth]{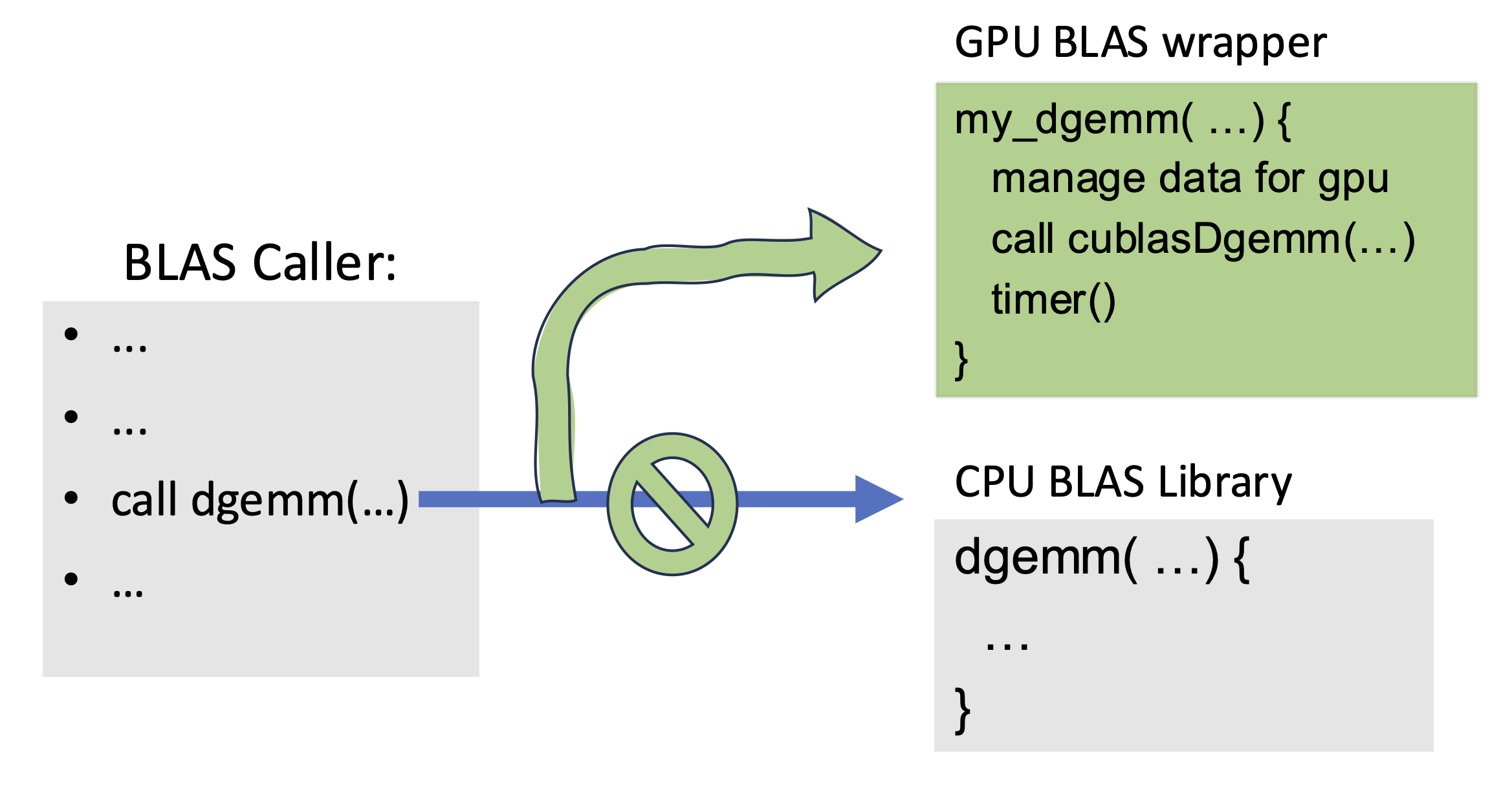}
    \caption{Workflow of Automatic BLAS Offload. \textnormal{The BLAS in the CPU binary is intercepted and replaced with a BLAS wrapper where GPU BLAS call is made and data is moved between CPU resident memory and GPU resident memory. }}
    \label{fig:workflow}
\end{figure}

A basic workflow of the automatic BLAS offload tool is illustrated in Figure \ref{fig:workflow} where a dgemm call in the caller code is intercepted and redirected to a BLAS wrapper that manages data movement and makes the GPU BLAS call.  
Conceptually, this workflow contains two tasks:
1) intercept BLAS symbols and replace them with a BLAS wrapper where GPU BLAS is called, 
2) manage data movement between CPU and GPU resident memories.  
In the following content, we will discuss how these tasks are implemented in SCILIB-Accel, and most critically how the data movement can be optimized using a novel data movement strategy inspired by the OpenMP First-Touch data placement policy. 


\subsection{Symbol Interception} 
Symbol interception is achieved via a trampoline-based Dynamic Binary Instrumentation (DBI) approach: 
a small piece of assembly code is inserted into the original function, enabling it to jump to a trampoline function. 
This trampoline function preserves the overwritten bytes by the extra jump instruction and executes customized code before returning to the original program. 

For automatic BLAS offload, we intercept BLAS calls where the trampoline function (BLAS wrapper function) has the same signature as the original function.
This design incurs minimal overhead, limited to the cost of two additional jump instructions. 
This mechanism finds extensive use in profilers, and here we use the PEAK\cite{peak} lightweight profiler as the DBI framework,  ensuring portability across various architectures including but not limited to x86 and ARM.
Inside the DBI framework, the FRIDA-GUM\cite{frida_gum} binary instrumentation backbend is chosen for simplicity and performance.   

The SCILIB-Accel automatic offload library can be attached to the user application
by preloading ($LD\_PRELOAD$) the SCILIB-Accel shared library file. 
The SCILIB-Accel initialization function is placed into the .init\_array section of Linux ELF to search and replace BLAS symbols along with other initialization tasks such as setting GPU memory pool, initializing cuBLAS, etc.  
Similarly, the SCILIB-Accel finalization function is inserted into the .fini\_array section of the ELF to collect statistics and handle clean up tasks. 

Note that the DBI approach applies to both dynamically and statically linked BLAS, 
while other tools like LIBSCI\_ACC\cite{cray-libsci-acc, cray-libsci-acc-slide} and NVBLAS\cite{nvidia-nvblas}, which works by resolving runtime library dependency, only work for dynamically linked BLAS.

As DBI is widely used in profilers, the DBI symbol interception in SCILIB-Accel can cause conflicts when doing profiling. 
To be profiler friendly, an implementation of SCILIB-Accel using dlsym() to dynamically resolve shared library symbols to intercept BLAS calls is also provided. 
This approach works by defining the  wrapper function to be the same name as the function to be intercepted, 
and by prepending the SCILIB-Accel library in $LD\_PRELOAD$, the symbols in the wrappers get used and the original function symbol can be obtained by looking up the next available symbol using dlsym(). 
This dlsym-based version has no issue being used with profilers, but can only intercept dynamically linked BLAS.

\subsection{Data Movement Strategies}

\begin{table*}[ht]
\centering
\centering
\caption{ OpenMP First-Touch vs Device First-Use } 
\vspace{-3mm}
\label{tab:dfu-policy}
\begin{tabular}{c|c}
\hline 
    \textbf { OpenMP First-Touch } & \textbf { Device First-Use } \\
\hline
   for dual socket CPU   &   for CPU-GPU superchip \\ 
   \includegraphics[scale=0.2]{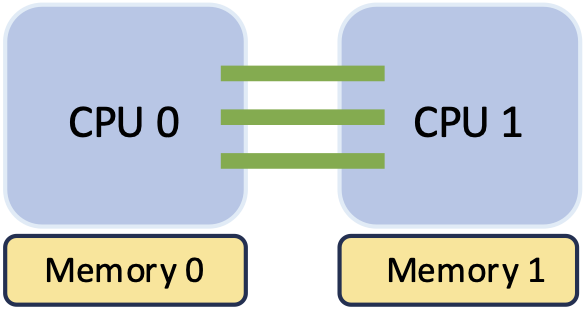} & \includegraphics[scale=0.2]{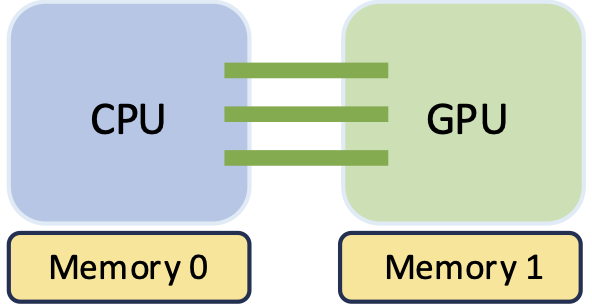} \\ 
    \hline
    allocate space on local memory of   & migrate data from CPU memory to device memory   \\ 
    OpenMP threads upon initialization &   upon first use by device (GPU) kernel \\
     \hline 
    assume remote memory access is trivial  & assume remote memory access is trivial  \\
    (e.g. CPU 0 accessing CPU 1’s memory) &  (e.g. CPU accessing GPU’s memory) \\ 

\hline
\end{tabular}
\end{table*}

Managing data movement is often the most critical part of GPU porting as data transfer speed has been a limiting factor.
It is even more so for developing an automatic offload tool, as the tool deals with pure CPU code that is totally untuned for GPU.
Here, we discuss three data movement strategies that utilize different features of Grace-Hopper so that this helps us better understand the challenge and opportunities for doing automatic offload on unified memory architecture. 

\subsubsection{Strategy 1, \textbf{\textit{ Mem-Copy}}} 

This is the most intuitive strategy and used by other tools. 
The pseudocode is listed below.
Upon interception of a BLAS call and redirection to a BLAS wrapper,
the input matrices are copied from host memory to GPU memory,  
and then resultant matrix is copied back after cuBLAS execution. 
This strategy works on all GPUs including the conventional PCIe-based cards without needing unified memory capability, 
but at the cost of frequent data movement.    
Such strategy can be effective for codes handling very large matrices where the compute time far exceeds the data transfer cost, but won't be useful for most codes that only runs small to medium sized matrices in practical use.  
This policy is studied here mostly for helping us understand the limitation of the conventional automatic BLAS offload approach in all other existing tools.


\begin{lstlisting}[language=C, 
  caption=Pseudocode: Mem-Copy data movement policy,
  basicstyle=\ttfamily,
  columns=fullflexible,
  literate=
    {host_A}{{{\textcolor{blue}{host\_A}}}}1
    {host_B}{{{\textcolor{blue}{host\_B}}}}1
    {host_C}{{{\textcolor{blue}{host\_C}}}}1    
    {device_A}{{{\textcolor{blue}{device\_A}}}}1
    {device_B}{{{\textcolor{blue}{device\_B}}}}1
    {device_C}{{{\textcolor{blue}{device\_C}}}}1
]
void my_dgemm(..., host_A, host_B, host_C, ...) {
...
GPU_Malloc(&device_A,size_A);
GPU_Malloc(&device_B,size_B);
GPU_Malloc(&device_C,size_C);

GPU_Memcpy(host_A to device_A,size_A);
GPU_Memcpy(host_B to device_B,size_B);

GPU_Dgemm(handle, TransA, TransB, m, n, k, 
            &alpha, device_A, lda, device_B, ldb, 
            &beta, device_C, ldc);
            
GPU_Memcpy(device_C to host_C,size_C); 
...
}
\end{lstlisting}

\subsubsection{Strategy 2, \textbf{\textit{counter-based migration}}}
    Since the CPU resident memory (LPDDR5X) and the GPU resident memory (HBM3) 
    are physically unified with cache-coherent NVLink C2C, 
    CPU matrix pointers can be passed directly to cuBLAS calls.  
    One can also use numactl to force all memory to be resident on the HBM, but given the poor bandwidth of CPU access HBM, a performance penalty is expected.  
    A new feature in Grace-Hopper designed to better serve the unified memory is the access counter on Hopper GPU, namely the CUDA runtime will automatically move memory pages from LPDDR5X to HBM3 based on remote memory accesses detected by the counter. 
    This counter-based migration mechanism can serve automatic offload when CPU resident matrices are passed to a GPU kernel.
    An example of a pseudocode is shown below. 
    

\begin{lstlisting}[language=C, 
  caption=Pseudocode: counter-based data migration policy,
  basicstyle=\ttfamily,
  columns=fullflexible,
  literate=
    {host_A}{{{\textcolor{blue}{host\_A}}}}1
    {host_B}{{{\textcolor{blue}{host\_B}}}}1
    {host_C}{{{\textcolor{blue}{host\_C}}}}1
]
void my_dgemm(..., host_A, host_B, host_C, ...) {
...
GPU_Dgemm(handle, TransA, TransB, m, n, k, 
      &alpha, host_A, lda, host_B, ldb, &beta, 
      host_C, ldc);
...
}
\end{lstlisting}

\subsubsection{ Strategy 3, \textbf{\textit Device First-Use } policy}

One alternative way to look at the Grace-Hopper superchip is that it operates as a heterogeneous dual-socket system, 
with one socket being a CPU and the other a GPU. 
Its NUMA configuration also mirrors that of a dual-socket CPU system, where CPU-resident memory is assigned to NUMA 0 and GPU-resident memory to NUMA 1.
The data management challenge here in CPU-GPU superchip is a reminiscence of the OpenMP First-Touch data placement policy used in CPU-CPU NUMA programming. 
While malloc calls can theoretically be intercepted to allocate memory directly for GPU access, these calls originate from the CPU binary, making it impractical to identify which memory regions will later be used by the GPU. 
As a result, implementing a GPU-first-touch policy is not feasible.
To address this, we propose a {\bf GPU First-Use policy}, 
where data is moved to the GPU the first time it is accessed by a CUDA kernel. 
More generally, this concept can be extended to a {\bf Device First-Use policy}, 
applicable to any accelerator device with cache-coherent memory access.
Table \ref{tab:dfu-policy} summarizes the features of Device First-Use and highlights its similarities to OpenMP First-Touch.

\begin{figure}[ht]
    \centering
    \includegraphics[width=\linewidth]{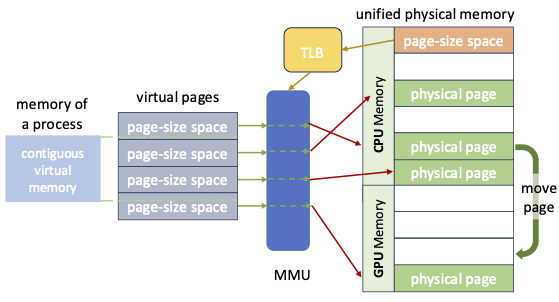}
    \caption{Virtual Memory and Physical Page Migration. \textnormal{Diagram showing translation between virtual and physical memory. The physical memory page could be moved without impacting virtual memory page or the viewpoint of memory from the executing binary side. }}
    \label{fig:vmem}
\end{figure}

Since the BLAS wrapper used to replace the CPU BLAS calls only knows the memory addresses of the matrices, 
implementing Device First-Use policy requires moving data from the CPU resident memory to device (GPU) resident memory 
without reallocation or disrupting to the virtual memory address used by the CPU binary.  
This can be achieved by relocating the physical memory page and updating the page table to remap the virtual memory to new physical memory locations. 
Figure \ref{fig:vmem} illustrates the working of  virtual memory in modern operation systems  and how a physical memory page can be dynamically reassigned.   

Remarkably, this complex process of physical page movement and remapping can be easily carried out using the Linux move\_page() system call. 
This system call allows specifying a group of pages to move along along with their target NUMA destination (NUMA 1 for GPU on Grace-Hopper), simplifying the implementation significantly.  
Pseudocode of this implementation is presented below. 

\begin{lstlisting}[language=C, 
  caption=Pseudocode: Device First-Use policy,
  basicstyle=\ttfamily,
  columns=fullflexible,
  literate=
    {host_A}{{{\textcolor{blue}{host\_A}}}}1
    {host_B}{{{\textcolor{blue}{host\_B}}}}1
    {host_C}{{{\textcolor{blue}{host\_C}}}}1
]
void my_dgemm(..., host_A, host_B, host_C, ...) {
...
if (host_A on NUMA 0) move_pages (host_A, to NUMA1) ;
if (host_B on NUMA 0) move_pages (host_B, to NUMA1) ;
if (host_C on NUMA 0) move_pages (host_C, to NUMA1) ;

GPU_Dgemm(handle, TransA, TransB, m, n, k, 
      &alpha, host_A, lda, host_B, ldb, &beta, 
      host_C, ldc);
... 
}
\end{lstlisting}

How does this policy improve data reusability for the GPU?
To answer this, let's examine how BLAS is typically used in scientific applications.
In most cases, scientific problems are not solved with a single BLAS call. 
Instead, they involve a sequence of BLAS operations, 
such as $C=A\times B$ followed by $E=D\times C$, and so on.
In these workflows, intermediate matrices (e.g., C) are frequently reused in subsequent operations.
Data reuse is also particularly common in block matrix multiplications, where each block of a matrix is multiplied by multiple blocks from another matrix. 
Additionally, many scientific codes adopts an iterative approach, such as the self-consistent field process to solve partial differential equations in quantum chemistry or physics, where the same matrices and memory pointers are re-used across all the iterations.  
All the data only need to be moved to the GPU once and can be re-used by subsequent iterations.  
All these common use cases conceptually justify the appropriateness of the Device First-Use policy. 

\subsection{Usage Instructions}
\label{sec:instruction}
The automatic BLAS offload wrappers are compiled and linked together 
as a shared library file (.so). 
All users need to do is to load the SCILIB-Accel library by $LD\_PRELOAD$ the .so file as shown below,  
and then run their CPU binary as normal.   

DBI version for general use: 
\begin{align*}
&\textit{export LD\_PRELOAD=/path/scilib-dbi.so} \\
&\textit{run your \textbf{CPU} binary}
\end{align*}

DLSYM-based version when using with profilers: 
\begin{align*}
&\textit{export LD\_PRELOAD=/path/scilib-dl.so} \\
&\textit{run your \textbf{CPU} binary}
\end{align*}

Several optional environmental variables can be set to tweak offload behaviors, including: 
\begin{enumerate}
    \item Data management strategies with Device First-Use as default.
    \item Minimum matrix size to be offloaded. \\
    If matrices are small, then the BLAS call will stay on CPU. The default threshold is $N_{avg}>500$ where $N_{avg}$ is the average matrix size, the definition of $N_{avg}$ is routine dependent. 
    For general matrix multiplication routines C=AxB, $N_{avg}=(MNK)^{1/3}$ where dimensions of matrices A, B and C are $M\times K$, $K\times N$ and $M\times N$.   
    The default threshold is a safe lower-bound based on preliminary dgemm testings on Grace-Hopper and can certainly be further fine-tuned for different kernels or precisions.  The optimal threshold is GPU-dependent. On Grace-Hopper, the default value is a conservative threshold obtained from performance evaluations and guarantees performance gain when running on GPU.  
     \item debug output levels.  
\end{enumerate}

\section{Performance Testings and Discussion}  
\label{sec:result} 
In this section, we perform application tests using production HPC codes. 
Since quantum chemistry/physics are known to heavily rely on BLAS operations, as linear algebra is the natural language of quantum mechanics, 
two codes, MuST and PARSEC, from this scientific domain are chosen. 
These codes are part of the Characteristic Science Application (CSA)\cite{csa} efforts funded by the National Scientific Foundation (NSF) as representative workloads for the Leadership Class Computing Facility. 

Different offload strategies are extensively tested to showcase the performance of Device First-Use data movement policy, 
and understand limitations of data movement approaches.  
In all test cases, the offload threshold is $N_{avg}>500$ (see Section \ref{sec:instruction} for details),
 which is proven to be appropriate for these applications on Grace-Hopper. 

It is important to emphasize that all performance comparisons presented here are based on an equal number of nodes: Grace-Grace nodes (two CPUs with 144 cores) and Grace-Hopper nodes (one CPU and one GPU). 
This approach ensures both simplicity and fairness in the analysis. 
Power and cost are also key factors in performance studies. 
The power consumption of a Grace-Hopper node is approximately twice that of a Grace-Grace node under full load. Additionally, for each node-hour on the Vista\cite{vista} supercomputer, the Service Unit (SU) charged to users for Grace-Hopper nodes is three times higher than that for Grace-Grace nodes. 
This charge rate reflects the costs associated with acquiring, maintaining, and supporting these different types of nodes.

\subsection{Test Environment} 

All tests were conducted on the Vista\cite{vista} supercomputer at the Texas Advanced Computing Center (TACC). 
The system comprises 560 Grace-Hopper (GH200) nodes and 180 Grace-Grace nodes, configured as follows:

\begin{itemize}
    \item \textbf{Grace-Hopper Nodes}: Each node is equipped with one 72-core Grace CPU (120 GB LPDDR5X on-board memory) and one H100 GPU (96 GB HBM3 memory). The Grace-Hopper superchip is power-capped at 900W.
    \item \textbf{Grace-Grace Nodes}: Each node features two 72-core Grace CPUs with a combined Thermal Design Power (TDP) of 500W. Each Grace chip has 120GB LPDDR5X memory integrated as well. 
\end{itemize}

Notably, the 120 GB Grace CPU model has approximately 30\% higher memory bandwidth compared to the 480 GB Grace model tested previously\cite{scilib-accel-short}.

The nodes are connected with Infiniband interconnects in non-blocking fat-tree topology:
\begin{itemize}
    \item Grace-Hopper nodes utilize full HDR (400 Gbps) configuration.
    \item Grace-Grace nodes are connected via split HDR (200 Gbps).
\end{itemize}

The software environment and configuration include:
\begin{itemize}
    \item \textbf{GPU Driver}: Version 560.35.03
    \item \textbf{CUDA}: Version 12.6
    \item \textbf{Infiniband Firmware}: Version 28.41.1000
    \item \textbf{NVHPC Compiler Suite}: Version 24.9 (latest at the time of testing)
    \item \textbf{MPI}: HPCX (based on OpenMPI 4.1.7a1) provided with the NVHPC compiler suite
    \item \textbf{OS}: Rocky Linux 9.3
    \item \textbf{Linux Kernel}: Version 5.14.0-362.24.1.el9\_3.aarch64+64k
\end{itemize}

For application testing, all CPU binaries were linked to the NVIDIA Performance Library (NVPL), which provides optimized BLAS, LAPACK, and ScaLAPACK routines. The SCILIB-Accel auto-offload tool utilized cuBLAS for GPU-accelerated operations. 
Both NVPL and cuBLAS were from the NVHPC 24.9 compiler suite.

\subsection{Application Test 1: MuST}  

\begin{table*}[ht]
    \caption{MuST: Performance on GPU vs CPU using 50 Nodes }\vspace{-3mm}
    \label{tab:must-test1}
    \centering
    \begin{tabular}{c|c|c|c|c}
        \hline \hline
        \multirow{2}{*}{Hardware} & \multirow{2}{*}{\textbf{Setup}} & \multirow{2}{*}{\textbf{\shortstack{Total\\ runtime (s)}} } & \multirow{2}{*}{\textbf{\shortstack{zgemm+\\ztrsm (s)}}} & \multirow{2}{*}{\textbf{\shortstack{Data \\movement (s)}}} \\ 
         & & & & \\ \hline 
        CPU: & \multirow{2}{*}{ CPU binary linked to NVPL} & \multirow{2}{*}{2318.4} & \multirow{2}{*}{2079.2} & \multirow{2}{*}{0} \\ 
        Grace-Grace &    &  & & \\  \hline
        \multirow{4}{*}{\shortstack{GPU:\\Grace-Hopper}} & native CUDA port & 1685 & N/A & N/A \\ \cline{2-5}
        &\shortstack{ auto offload: Mem-Copy} & 1098 & 439.8 & 291.7 \\ \cline{2-5}
        &\shortstack{auto offload: counter-based migration} & 858 & 616.0 & included in BLAS \\ \cline{2-5}
        &\shortstack{auto offload: \textbf{Device First-Use}} & \textbf{824} & \textbf{580.0} & \textbf{4.8}$^\dagger$\\ \hline\hline 
\multicolumn{5}{l}{\scriptsize $^\dagger$ Matrices migrated to GPU resident memory are reused 780 times. } \\
    \end{tabular}
\end{table*}

\begin{table*}[ht] 
\caption{ MuST: Strong Scaling Performance on CPU vs GPU } \vspace{-3mm}
    \centering
    \label{tab:must-test2}
    \begin{tabular}{c|c|c|c|c}
        \hline\hline
      \multirow{3}{*}{\textbf{\shortstack{Node Count}}}  & \multicolumn{3}{c|}{Total runtime (s) } & \multirow{3}{*}{\textbf{\shortstack{Best \\GPU/CPU\\ speedup}}} \\ 
        \cline{2-4} 
         & \multirow{2}{*}{\textbf{\shortstack{CPU \\(Grace-Grace)}}} & \textbf{GPU: Native } & \textbf{GPU, auto offload:} &  \\
         &    & \textbf{ CUDA port} & \textbf{Device First-Use} &  \\
        \hline
        25  & 4598.1 & 3223.3 & 1550.9 & 3.0x \\
        50  & 2318.4 & 1685.2 & 823.8  & 2.8x \\
        75  & 1842.6 & 1244.7 & 623.1  & 3.0x \\
         100 & 1192.2  & 903.9  & 446.8  & 2.7x \\
        150 & 947.0 & 673.6  & 357.5  & 2.6x \\
        200 & N/A  $^\dagger$    & 493.9  & 253.3  & N/A $^\dagger$  \\
        \hline\hline
      \multicolumn{5}{l}{\scriptsize $^\dagger$ Not enough CPU nodes available. } \\
    \end{tabular}
\end{table*}

\begin{figure}[ht]
    \centering
    \includegraphics[width=0.95\linewidth]{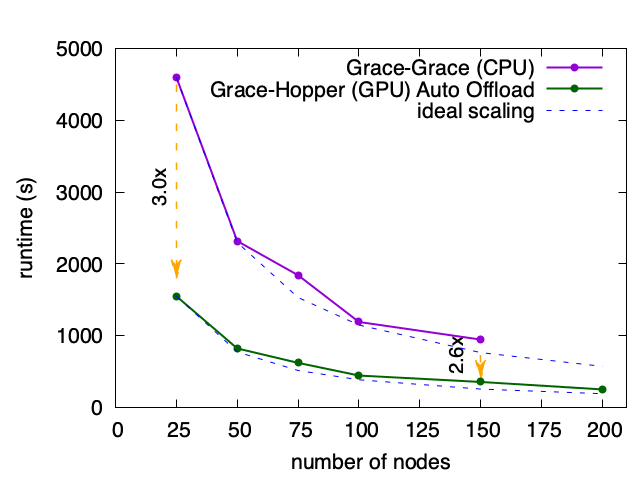}
    \caption{MuST: Strong Scaling Test for CPU Run and Automatic GPU Offload Run. \textnormal{Energy calculation of the 5600-atom CoCrFeMnNi alloy system is tested on on various number of Grace-Grace nodes (2 Grace CPUs per node) and Grace-Hopper nodes (1 CPU and 1 GPU per node). Speedup of automatic BLAS offload on GPU nodes is up to 3x comparing to the same number of CPU nodes, breaking even with the cost ratio of node-hour charged for the Vista\cite{vista} system at TACC.}}
    \label{fig:must-scaling}
\end{figure}

MuST (Multiple Scattering Theory)\cite{must1,must2} is a package designed to perform electronic structure calculations, it solves the Kohn-Sham equation by solving the Green's function.
The LSMS calculation method is designed for large systems with linear scalability to the system size. 
This method is won the 2009 Gordon Bell prize \cite{gordonbell09}.
The code has a heavy dependency on BLAS operations, mostly zgemm and ztrsm, which often exceed 80\% of the runtime on CPU. A major portion of the BLAS calls are from the LAPACK routines zgetrs and zgetrf.  
Most matrices are squared or near-square shaped. 
MuST natively supports GPU offload through a CUDA implementation and offloads the matrix inverse onto a GPU by calling cuSOLVER.

The test workload calculates the energy of a CoCrFeMnNi supercell alloy using the LSMS method. 
Total number of atoms in the supercell is 5600 and the concentration of each element is identical. 
The number of energy grid is 32. 
The calculation is limited to 3 self-consistent field (SCF) steps in order to reduce benchmark cost.

MuST is thoroughly tested at large scale on CPU,  and on GPU through both the automatic offload and the native CUDA port.   
Table \ref{tab:must-test1} summarizes the performance comparison of different run strategies on 50 Grace-Grace CPU nodes or 50 Grace-Hopper nodes.
As mentioned before, the code strongly relies on BLAS operations. In this particular test workload, the two major BLAS routines zgemm and ztrsm consumes about 2080s out of the total 2318s runtime.  
Using the native CUDA port from the developers, about 1.4x speedup is achieved comparing to the CPU-only execution. 
Surprisingly, all auto offload strategies are faster than the native CUDA code, illustrating the complexity and challenge of CUDA programming, the developers will need spend substantial more amount porting efforts to polish their CUDA code.  
With the most basic data management method that copies (cudaMemcpy) matrices to/from GPU for every cuBLAS call, 
the total runtime is reduced to 1098s, but 292s are spent in just moving the data around.  
This reaffirms that optimizing data movement is still critical on Grace-Hopper even with the fast NVLink-C2C interconnect. 
The counter-based migration approach works okay, as the total runtime is better than doing frequent cudaMemcpy. Note that the page migration time is included in the BLAS call time as the counter-based does the data movement automatically behind the scene with GPU kernel.  
The novel Device First-Use policy is substantially better than other approaches, since the total runtime is reduced to 824s, about 2.8x faster than the CPU run, and the total data movement time is reduced to just 4.8s.  
More analysis shows that under  Device First-Use policy, each matrix that gets migrated to the GPU resident memory gets reused 780 times by subsequent BLAS calls. 
Such high level of data reusage is the key factor of the good performance that is achieved here.    
Also note that the BLAS (zgemm and ztrsm)  time in Device First-Use is much longer than the corresponding time in Mem-Copy policy.
This is due to a performance issue of CUDA kernel accessing GPU memory allocated by system malloc, more details are discussed in Section \ref{sec:align-issue}. If this performance issue can be resolved by NVIDIA, the performance of automatic offload can be further improved. 

Large scale strong scaling tests were also performed for this code.   
Table \ref{tab:must-test2} shows the runtime of CPU run, GPU run with native CUDA code, and automatic offload using Device First-Use policy. 
Scaling range goes from 25 nodes to 200 nodes.   
Throughout the tests, the automatic offload approach is consistently 2x faster than the native CUDA code, and up to 3x faster than the CPU run, breaking even with the extra node-hour charge rate for GPU, so users running on GPU not only get faster time-to-solution but also more efficiency in terms of energy consumption and cost.   
The strong scaling data is visualized in Figure \ref{fig:must-scaling}, both the CPU run and automatic offload GPU run have excellent scability, reaching very close to linear scaling at this wide test range.

\subsection{Application Test 2: PARSEC} 
\label{sec:parsec} 

\definecolor{intelblue}{RGB}{0, 199, 253} 
\definecolor{nvidiagreen}{RGB}{118, 185, 0}

\begin{table*}[ht!]
    \caption{PARSEC: Performance on GPU vs CPU on single node }\vspace{-3mm}
    \label{tab:parsec-test}
    \centering
    \begin{tabular}{c|c|c|c|c}
        \hline\hline
        \multirow{2}{*}{\textbf{Hardware}} & \multirow{2}{*}{\textbf{Setup}} & \multirow{2}{*}{\textbf{\shortstack{Total\\ runtime (s)}}} & \multirow{2}{*}{\textbf{\shortstack{dgemm (s)}}} & \multirow{2}{*}{\textbf{\shortstack{Data \\movement (s)}}} \\ 
        & & & & \\ \hline
        CPU: &  \multirow{2}{*}{CPU binary linked to NVPL} & \multirow{2}{*}{ 415.1 } & \multirow{2}{*}{270.1} & \multirow{2}{*}{0} \\  
        Grace-Grace &   &  & & \\  \hline
      \multirow{3}{*}{\shortstack{GPU:\\Grace-Hopper}}  &\shortstack{ auto offload: Mem-Copy} & 425.7 & 12.4 & 220.7 \\ \cline{2-5}
        &\shortstack{auto offload: counter-based migration} & 470.0 & 234.0 & included in BLAS \\ \cline{2-5}
        &\shortstack{auto offload: \textbf{Device First-Use}} & \textbf{220.3} & \textbf{29.1} & \textbf{1.3}$^\dagger$ \\ \hline\hline
    \multicolumn{5}{l}{\scriptsize $^\dagger$ Matrices migrated to GPU resident memory are reused 570 times. } \\
    \end{tabular}
\end{table*}

PARSEC (Pseudopotential Algorithm for Real-Space Electronic Calculations)\cite{parsec1,parsec2} is a package designed to perform Density Functional Theory (DFT) calculations of solids and molecules.
It solves the Kohn–Sham equations directly in real space, avoiding the use of explicit basis sets.  
Our benchmark case calculates energy of a Silicon nanocrystal $Si_{1947}H_{604}$, boundary sphere radius is set to 50 bohr, grid spacing is 0.9 bohr, the calculation is limited to two self-consistent field steps to reduce the benchmark cost, but the performance characteristics of a fully converged calculation are identical. 

 Historically, PARSEC has been a CPU-only code, relying heavily on ScaLAPACK.
 In typical use cases, dgemm calls  from ScaLAPACK can account for over 50\% of the runtime.
 With the help of SCILIB-Accel automatic offload,  
 PARSEC runs on GPU for the first time with good performance.  
Tests were conducted to evaluate all offload strategies alongside a CPU-only baseline. All tests were performed on a single node: Grace-Grace for CPU runs and Grace-Hopper for GPU runs. The results are summarized in Table \ref{tab:parsec-test}.

In these tests, the Mem-Copy data policy resulted in runtimes slower than dual-CPU execution. 
The cudaMemcpy operations consumed 220 seconds, accounting for more than 50\% of the total runtime — a significantly higher proportion than the one observed in the MuST test case.
This is due to the fact that most matrices used in PARSEC are long skinny matrices rather than squared shape ones.
For example, a common dgemm input in PARSEC is \textit{transA='T', transB='N', M=32, N=2400, K=93536}, this extreme skinny shapes make the total byte size of the matrices much bigger than if square matrices are used in a calculation with equivalent computational workload. 
This outcome again demonstrates that the conventional data movement strategy for automatic offload is  impractical even on Grace-Hoper where NVLink-C2C transfer rate is 450 GB/s per direction, 
not to mention the PCIe based cards where PCIe Gen5 x16 can only do 64 GB/s.  
The counter-based data migration strategy performs even worse due to incompetent migration algorithm, 
see complete discussions of NVIDIA's migration issues later in Section \ref{sec:counter-issue}. 

Finally, the Device First-Use policy is able to efficiently manage the data movement, enabling significant performance gains.
The total runtime is nearly 2x faster than the CPU run, with the dgemm component achieving nearly 10x speedup compared to the CPU run. 
The data transfer overhead is minimal, totaling just 1.3 seconds.  
The speedup is able to offset the extra cost from power hungry GPU. 
Again, data reuse is counted and for every matrix that is migrated to the GPU resident memory, it is reused on average 570 times by subsequent dgemm calls. 

\subsection{Performance Issues with Grace-Hopper Relevant to Auto Offload} 
In this part of the paper, we discuss a few performance issues observed in the above application test, 
these issues reflect the immaturity of the software or hardware design in the current Grace-Hopper system, 
and could be fixed in the future to further improve performance and usability of automatic offload. 

\subsubsection{ \textit{\textbf {Counter-based page migration}}} 
\label{sec:counter-issue} 

\begin{table}[ht]
\centering\captionsetup{justification=centering}
\caption{Behavior of CUDA Counter-based Data Migration: by \texttt{cublasDgemm} using host malloc pointers} \vspace{-3mm}
\label{tab:dgemm-migration}
\begin{tabular}{c|ccc|ccc}
\hline
\hline
Matrix & \multicolumn{3}{c|}{Matrix Sizes} & \multicolumn{3}{c}{CPU->GPU} \\
Dimensions    &  \multicolumn{3}{c|} { (in MB) } & \multicolumn{3}{c}{migrated$^\dagger$}\\
\hline 
 (M, N, K) & A & B & C &  A & B & C \\
\hline
(1000, 1000, 1000)    &  8 & 8 & 8 & yes & yes & yes \\
(5000, 5000, 5000)    &  200  & 200 & 200 & yes? & yes? &  no \\
(20000, 20000, 20000) &  3200 & 3200 & 3200 & yes & no & no  \\
(32, 2400, 93536)     & 24 & 1796 & 0.6 &  yes &  no & no \\
\hline
\hline 
\multicolumn{7}{l}{\scriptsize $^\dagger$ Result with question mark (?) indicates inconsistent run-to-run behavior. } \\
\end{tabular}
\end{table}

The Hopper GPU has an access counter that monitors remote memory access, 
and migrates memory pages from CPU resident memory to GPU resident memory. 
Due to the lack of access counter on the Grace CPU, data on GPU will not be migrated back to CPU. 
The details of the migration criteria are unknown.  

Since the data migrated to GPU will not be migrated back, 
automatic offload with counter-based migration should work very similar to manually implemented Device First-Use policy, 
but we have already seen the slow performance of the counter-based migration in application tests. 
Here we present a few simple dgemm test cases with different matrix sizes, 
and provide a deeper understanding of the issues with the counter-based migration.   
In the following test, matrices A(M,K), B(K,N) and C(M,N) are allocated by malloc and initialized on CPU resident memory, 
then multiplication $C=A\times B$ is performed at least 5 times by passing these matrices to cublasDgemm, 
so that the access-counter should allow the matrices to be migrated to GPU resident memory.  
The NUMA locations of the matrices are reported after each cublasDgemm call, and runtime for each call is reported.
From the NUMA location, we can infer whether the data is on the CPU resident memory (NUMA 0) or GPU resident memory (NUMA 1). 
Four sets of inputs with different matrix sizes and shapes are tested, all of which use FP64 precision. 

When M=N=K=1000 are used, the sizes of A, B and C are all 8.0MB. 
All of them are successfully migrated to the GPU resident memory upon the first cublasDgemm call. 

When the matrix dimensions become M=N=K=5000, the sizes of all the matrices are 200.0MB, 
and the migration becomes unstable and inconsistent from run-to-run. 
The tests were run many times. In most cases, only matrices A and B are migrated to GPU in the first cublasDgemm call, while C stays on the CPU no matter how many more cublasDgemm call iterations are added. 
Occasionally, matrices A and B stays on the CPU side in the first cycle, and only get migrated to GPU after the second call. 
In all runs, matrix C is never migrated to GPU.   

If the square matrix dimension is increased to M=N=K=20000, matrix size becomes 3200.0MB each, 
only matrix A is migrated to GPU, 
while B and C are always on the CPU memory no matter how many more cublasDgemm cycles are added.  

The case gets strange for non-square matrix dimensions.  
When M=32, N=2400 and K=93536, a matrix size commonly used in the PARSEC workload, 
the matrix size becomes 24.0MB for A, 1795.9MB for B, and 0.6MB for C.  
Throughout our test, only matrix A gets migrated to GPU, while B and C are always on the CPU. 
This is very counter-intuitive as one would expect the bigger matrix B should at least be moved, as it generates the largest volume of remote memory accesses, but NVIDIA's counter-based migration algorithm fails to move it. 

From these tests, we can see that NVIDIA's algorithm tends to migrate smaller data to GPU. 
The reason could be that the algorithm only makes the decision based on a single CUDA kernel call,
i.e. moving the big data is costly comparing to that single CUDA kernel runtime, so it decides not to migrate 
disregarding the fact that the big data gets accessed by the GPU many more times later on.   
The current counter-based migration is thus unpredictable and inconsistent, 
and NVIDIA should provide a simple way to disable it by users instead of requiring to unload a kernel module by root.

\subsubsection{\textit{\textbf {Impact of page sizes on counter-based migration}}}
\label{issue-page}

\begin{table}[ht]
\centering\captionsetup{justification=centering}
\caption{DGEMM Runtime with Unified Memory} \vspace{-3mm}
\label{tab:dgemm-pagesize}
\begin{tabular}{ c | c | c c}
\hline\hline 
\multirow{2}{*}{Page Size} & \multirow{2}{*}{Memory Type} & CPU (72C) & GPU    \\
 & & (dgemm) & (cublasDgemm) \\
\hline\hline 
 \multicolumn{4}{c}{ \hspace{0.5cm} Workload: M=2000, N=2000, K=2000; 96MB total} \\
\hline
\multirow{2}{*}{4KB} & LPDDR5X & 5.1 ms & 9.0 ms \\
                     & HBM3 & 5.3 ms & 0.37 ms  \\
\hline 
\multirow{2}{*}{64KB} & LPDDR5X & 5.1 ms & N/A$^\dagger$ \\
                     & HBM3 & 10.0 ms & 0.39 ms  \\
\hline\hline 
  \multicolumn{4}{c}{ \hspace{0.5cm} Workload: M=32, N=2400, K=93536; 1820MB total} \\
\hline
\multirow{2}{*}{4KB} & LPDDR5X & 10.9 ms & 18.1 ms \\
                     & HBM3 & 15.5 ms & 0.95 ms \\   
\hline                      
\multirow{2}{*}{64KB} & LPDDR5X & 15.8 ms & N/A$^\dagger$ \\
                     & HBM3 & 23.2 ms & 0.94 ms \\
                     
\hline \hline 
\multicolumn{4}{l}{\scriptsize $^\dagger$ Part of the data gets migrated to HBM3 and can't do full LPDDR5X run. } \\
\end{tabular}
\end{table}

The Grace-Hopper platform, similar to all other ARM-based platforms, supports two base page sizes 4KB and 64KB. 
Most of the NVIDIA's internal tests are done on 64KB page size which is also the recommended page size. 
At TACC,  both page sizes are setup for testings, and performance issues with the 64KB page size are revealed by comparing the test results. 

Here we again run simple dgemm tests to understand the performance and issues. 
Notice that the aforementioned counter-based page migration mechanism does not work with the 4KB page size, 
so one can measure the performance of GPU kernel running on LPDDR5X, 
while this cannot be done when page size is 64KB as matrices are partially migrated to GPU as explained in Section \ref{sec:counter-issue}. 

The test results are summarized in Table \ref{tab:dgemm-pagesize}.  
It can be seen that CPU accessing HBM3 memory is substantially slower under 64KB page than 4KB page for both problem sizes.  
There is a substantial difference even for CPU accessing LPDDR5X for the second workload,
runtime with 64KB page is 15.8ms, much slower than the 10.9ms under 4KB page.

\subsubsection{\textit{\textbf {Sensitivity of page alignment for GPU accessing system allocated HBM}}} 
\label{sec:align-issue}



Application tests reveal that BLAS runtime performance under the Device First-Use policy is noticeably slower compared to BLAS operating on cudaMalloc memory used in the Mem-Copy policy. Further investigation indicates that CUDA kernel performance is suboptimal when operating on system-allocated HBM3 via malloc unless the matrices are aligned to the page.
Table \ref{tab:dgemm-align} highlights the performance differences. The tests use 64KB page size.  All matrices are allocated using malloc and pinned to HBM3 with the command \textit{numactl -m 1 ./exe}. 
When matrices are page-aligned, the performance of cublasDgemm improves by nearly 50\% compared to cases where the data is not aligned. 
This improvement is particularly pronounced for memory-bandwidth-bound kernels.
In the page-aligned case, the performance on HBM3 allocated by malloc matches that of the same CUDA kernel executing on cudaMalloc memory. 
The reason of such behavior is unknown, and it partially defeats the advantage of unified memory architecture.

\begin{table}[ht]
\centering\captionsetup{justification=centering}
\caption{Impact of Memory Alignment \\ on cublasDgemm Performance$^\dagger$} \vspace{-3mm}
\label{tab:dgemm-align}
\begin{tabular}{ c  | c c}
\hline 
problem size &  unaligned & aligned    \\
\hline 
 M=2000, N=2000, K=2000 &  0.39 ms & 0.29 ms \\
\hline
M=32, N=2400, K=93536  &   0.94 ms & 0.64 ms \\
\hline  
\multicolumn{3}{l}{\scriptsize $^\dagger$ Tests were performed on 64KB page size, all memory allocations are by malloc. } \\
\end{tabular}
\end{table}

\section{Conclusion and Future Work}
\label{sec:conclusion}

In this paper,  a proof-of-concept BLAS auto-offload prototype tool is further optimized and extended to all level-3 BLAS operations. 
 To overcome the data transfer bottleneck between CPU and GPU, an OpenMP first-touch type of data management strategy, namely the GPU first-use policy, is proposed and discussed in details. 
 The new policy minimizes the data transfer between CPU and GPU in practical BLAS-heavy scientific computing codes, such as the quantum chemistry and quantum physics applications. Performance tests for several such codes show production-level GPU performance. 
 As all performance evaluations were performed on the NVIDIA Grace-Hopper architecture, 
 limitations and advantages of the memory subsystem in Grace-Hopper are outlined. 
 The proposed GPU first-use data management policy is applicable not only to the architecture tested here but can be universally applied for any CPU-GPU architecture that allows cache-coherent access of the GPU memory from CPU. 
This tool enables domain scientists to rapidly adopt modern GPUs in complex or legacy, BLAS-heavy codes to speed up scientific discoveries, and can also be used to quickly assess the potential benefits of GPU acceleration before a formal effort of porting.
The work in progress is to support auto BLAS offload on AMD GPUs. Additionally,  FFTW auto-offload is under exploration. 

\begin{acks}

The author J. Li thanks Robert Henschel for teaching him about the automatic offload capability in Cray libsci library and OpenMP First-Touch about seven years ago when Li switched his career from quantum chemistry to HPC.  

This work is supported by the National Science Foundation through awards OAC-2402542, OAC-1854828, and OAC-2139536. 

\end{acks}

\bibliographystyle{ACM-Reference-Format}
\bibliography{reference}


\begin{thebibliography}{16}


\ifx \showCODEN    \undefined \def \showCODEN     #1{\unskip}     \fi
\ifx \showDOI      \undefined \def \showDOI       #1{#1}\fi
\ifx \showISBNx    \undefined \def \showISBNx     #1{\unskip}     \fi
\ifx \showISBNxiii \undefined \def \showISBNxiii  #1{\unskip}     \fi
\ifx \showISSN     \undefined \def \showISSN      #1{\unskip}     \fi
\ifx \showLCCN     \undefined \def \showLCCN      #1{\unskip}     \fi
\ifx \shownote     \undefined \def \shownote      #1{#1}          \fi
\ifx \showarticletitle \undefined \def \showarticletitle #1{#1}   \fi
\ifx \showURL      \undefined \def \showURL       {\relax}        \fi
\providecommand\bibfield[2]{#2}
\providecommand\bibinfo[2]{#2}
\providecommand\natexlab[1]{#1}
\providecommand\showeprint[2][]{arXiv:#2}

\bibitem[Chelikowsky et~al\mbox{.}(1994)]%
        {parsec2}
\bibfield{author}{\bibinfo{person}{James~R. Chelikowsky}, \bibinfo{person}{N. Troullier}, {and} \bibinfo{person}{Y. Saad}.} \bibinfo{year}{1994}\natexlab{}.
\newblock \showarticletitle{Finite-difference-pseudopotential method: Electronic structure calculations without a basis}.
\newblock \bibinfo{journal}{\emph{Phys. Rev. Lett.}}  \bibinfo{volume}{72} (\bibinfo{year}{1994}), \bibinfo{pages}{1240--1243}.
\newblock
Issue 8.
\urldef\tempurl%
\url{https://doi.org/10.1103/PhysRevLett.72.1240}
\showDOI{\tempurl}


\bibitem[Eisenbach et~al\mbox{.}(2017)]%
        {must2}
\bibfield{author}{\bibinfo{person}{Markus Eisenbach}, \bibinfo{person}{Jeff Larkin}, \bibinfo{person}{Justin Lutjens}, \bibinfo{person}{Steven Rennich}, {and} \bibinfo{person}{James~H. Rogers}.} \bibinfo{year}{2017}\natexlab{}.
\newblock \showarticletitle{GPU acceleration of the Locally Selfconsistent Multiple Scattering code for first principles calculation of the ground state and statistical physics of materials}.
\newblock \bibinfo{journal}{\emph{Computer Physics Communications}}  \bibinfo{volume}{211} (\bibinfo{year}{2017}), \bibinfo{pages}{2--7}.
\newblock
\showISSN{0010-4655}
\urldef\tempurl%
\url{https://doi.org/10.1016/j.cpc.2016.07.013}
\showDOI{\tempurl}


\bibitem[Eisenbach et~al\mbox{.}(2009)]%
        {gordonbell09}
\bibfield{author}{\bibinfo{person}{M. Eisenbach}, \bibinfo{person}{C.-G. Zhou}, \bibinfo{person}{D.~M. Nicholson}, \bibinfo{person}{G. Brown}, \bibinfo{person}{J. Larkin}, {and} \bibinfo{person}{T.~C. Schulthess}.} \bibinfo{year}{2009}\natexlab{}.
\newblock \showarticletitle{A scalable method for ab initio computation of free energies in nanoscale systems}. In \bibinfo{booktitle}{\emph{Proceedings of the Conference on High Performance Computing Networking, Storage and Analysis}} (Portland, Oregon) \emph{(\bibinfo{series}{SC '09})}. \bibinfo{publisher}{Association for Computing Machinery}, \bibinfo{address}{New York, NY, USA}, Article \bibinfo{articleno}{64}, \bibinfo{numpages}{8}~pages.
\newblock
\showISBNx{9781605587448}
\urldef\tempurl%
\url{https://doi.org/10.1145/1654059.1654125}
\showDOI{\tempurl}


\bibitem[Foundation(2021)]%
        {csa}
\bibfield{author}{\bibinfo{person}{National~Science Foundation}.} \bibinfo{year}{2021}\natexlab{}.
\newblock \bibinfo{title}{Characteristic Science Applications for the Leadership Class Computing Facility}.
\newblock \bibinfo{howpublished}{\url{https://www.nsf.gov/awardsearch/showAward?AWD_ID=2139536&HistoricalAwards=false}}.
\newblock


\bibitem[Frida(nd)]%
        {frida_gum}
\bibfield{author}{\bibinfo{person}{Frida}.} \bibinfo{year}{n.d.}\natexlab{}.
\newblock \bibinfo{title}{Frida Gum}.
\newblock \bibinfo{howpublished}{\url{https://github.com/frida/frida-gum}}.
\newblock
\newblock
\shownote{Accessed: 2024-12-27}.


\bibitem[{Hewlett Packard Enterprise}({[n.\,d.]})]%
        {cray-libsci-acc}
\bibfield{author}{\bibinfo{person}{{Hewlett Packard Enterprise}}.} \bibinfo{year}{[n.\,d.]}\natexlab{}.
\newblock \bibinfo{booktitle}{\emph{HPE Cray Programming Environment documentation}}.
\newblock
\urldef\tempurl%
\url{https://h41374.www4.hpe.com/docs/csml/cray_libsci_acc.html}
\showURL{%
\tempurl}


\bibitem[{IBM}({[n.\,d.]})]%
        {ibm-essl-offload}
\bibfield{author}{\bibinfo{person}{{IBM}}.} \bibinfo{year}{[n.\,d.]}\natexlab{}.
\newblock \bibinfo{booktitle}{\emph{IBM Engineering and Scientific Subroutine Library for Linux on POWER}}.
\newblock
\urldef\tempurl%
\url{https://www.ibm.com/docs/en/SSFHY8_6.1/reference/essl_reference_pdf.pdf}
\showURL{%
\tempurl}


\bibitem[Kronik et~al\mbox{.}(2006)]%
        {parsec1}
\bibfield{author}{\bibinfo{person}{Leeor Kronik}, \bibinfo{person}{Adi Makmal}, \bibinfo{person}{Murilo~L. Tiago}, \bibinfo{person}{M.~M.~G. Alemany}, \bibinfo{person}{Manish Jain}, \bibinfo{person}{Xiangyang Huang}, \bibinfo{person}{Yousef Saad}, {and} \bibinfo{person}{James~R. Chelikowsky}.} \bibinfo{year}{2006}\natexlab{}.
\newblock \showarticletitle{PARSEC – the pseudopotential algorithm for real-space electronic structure calculations: recent advances and novel applications to nano-structures}.
\newblock \bibinfo{journal}{\emph{physica status solidi (b)}} \bibinfo{volume}{243}, \bibinfo{number}{5} (\bibinfo{year}{2006}), \bibinfo{pages}{1063--1079}.
\newblock
\urldef\tempurl%
\url{https://doi.org/10.1002/pssb.200541463}
\showDOI{\tempurl}


\bibitem[Li and Wang(2024)]%
        {scilib-accel-code}
\bibfield{author}{\bibinfo{person}{Junjie Li} {and} \bibinfo{person}{Yinzhi Wang}.} \bibinfo{year}{2024}\natexlab{}.
\newblock \bibinfo{booktitle}{\emph{SCILIB-accel: automatic BLAS offload tool}}.
\newblock
\urldef\tempurl%
\url{{https://github.com/nicejunjie/scilib-accel}}
\showURL{%
\tempurl}


\bibitem[Li et~al\mbox{.}(2024)]%
        {scilib-accel-short}
\bibfield{author}{\bibinfo{person}{Junjie Li}, \bibinfo{person}{Yinzhi Wang}, \bibinfo{person}{Xiao Liang}, {and} \bibinfo{person}{Hang Liu}.} \bibinfo{year}{2024}\natexlab{}.
\newblock \showarticletitle{Automatic BLAS Offloading on Unified Memory Architecture: A Study on NVIDIA Grace-Hopper}. In \bibinfo{booktitle}{\emph{Practice and Experience in Advanced Research Computing 2024: Human Powered Computing}} (Providence, RI, USA) \emph{(\bibinfo{series}{PEARC '24})}. \bibinfo{publisher}{Association for Computing Machinery}, \bibinfo{address}{New York, NY, USA}, Article \bibinfo{articleno}{47}, \bibinfo{numpages}{5}~pages.
\newblock
\showISBNx{9798400704192}
\urldef\tempurl%
\url{https://doi.org/10.1145/3626203.3670561}
\showDOI{\tempurl}


\bibitem[NVIDIA(2023)]%
        {gh-whitepaper}
\bibfield{author}{\bibinfo{person}{NVIDIA}.} \bibinfo{year}{2023}\natexlab{}.
\newblock \bibinfo{title}{NVIDIA GH200 Grace Hopper Superchip Architecture}.  (\bibinfo{year}{2023}).
\newblock
\urldef\tempurl%
\url{https://resources.nvidia.com/en-us-grace-cpu/nvidia-grace-hopper}
\showURL{%
\tempurl}


\bibitem[{NVIDIA}(2024)]%
        {nvidia-nvblas}
\bibfield{author}{\bibinfo{person}{{NVIDIA}}.} \bibinfo{year}{2024}\natexlab{}.
\newblock \bibinfo{booktitle}{\emph{NVBLAS documentation}}.
\newblock
\urldef\tempurl%
\url{https://docs.nvidia.com/cuda/nvblas}
\showURL{%
\tempurl}


\bibitem[Poxon(2013)]%
        {cray-libsci-acc-slide}
\bibfield{author}{\bibinfo{person}{Heidi Poxon}.} \bibinfo{year}{2013}\natexlab{}.
\newblock \bibinfo{title}{Introduction to the Cray Accelerated Scientific Libraries}.  (\bibinfo{year}{2013}).
\newblock
\urldef\tempurl%
\url{https://www.olcf.ornl.gov/wp-content/uploads/2013/01/Scientific_Libs.pdf}
\showURL{%
\tempurl}


\bibitem[{Texas Advanced Computing Center}(2024)]%
        {vista}
\bibfield{author}{\bibinfo{person}{{Texas Advanced Computing Center}}.} \bibinfo{year}{2024}\natexlab{}.
\newblock \bibinfo{title}{VISTA System at TACC}.
\newblock
\newblock
\urldef\tempurl%
\url{https://tacc.utexas.edu/systems/vista}
\showURL{%
\tempurl}


\bibitem[Wang and Li(2023)]%
        {peak}
\bibfield{author}{\bibinfo{person}{Yinzhi Wang} {and} \bibinfo{person}{Junjie Li}.} \bibinfo{year}{2023}\natexlab{}.
\newblock \showarticletitle{{PEAK}: a {Light}-{Weight} {Profiler} for {HPC} {Systems}}. In \bibinfo{booktitle}{\emph{Proceedings of the {SC} '23 {Workshops} of {The} {International} {Conference} on {High} {Performance} {Computing}, {Network}, {Storage}, and {Analysis}}} \emph{(\bibinfo{series}{{SC}-{W} '23})}. \bibinfo{publisher}{Association for Computing Machinery}, \bibinfo{address}{New York, NY, USA}, \bibinfo{pages}{677--680}.
\newblock
\showISBNx{9798400707858}
\urldef\tempurl%
\url{https://doi.org/10.1145/3624062.3624143}
\showDOI{\tempurl}


\bibitem[Wang et~al\mbox{.}(1995)]%
        {must1}
\bibfield{author}{\bibinfo{person}{Yang Wang}, \bibinfo{person}{G.~M. Stocks}, \bibinfo{person}{W.~A. Shelton}, \bibinfo{person}{D.~M.~C. Nicholson}, \bibinfo{person}{Z. Szotek}, {and} \bibinfo{person}{W.~M. Temmerman}.} \bibinfo{year}{1995}\natexlab{}.
\newblock \showarticletitle{Order-N Multiple Scattering Approach to Electronic Structure Calculations}.
\newblock \bibinfo{journal}{\emph{Phys. Rev. Lett.}}  \bibinfo{volume}{75} (\bibinfo{date}{Oct} \bibinfo{year}{1995}), \bibinfo{pages}{2867--2870}.
\newblock
Issue 15.
\urldef\tempurl%
\url{https://doi.org/10.1103/PhysRevLett.75.2867}
\showDOI{\tempurl}


\end{thebibliography}



 \end{document}